\documentclass[sigplan,nonacm]{acmart}
\pdfoutput=1
\renewcommand\footnotetextcopyrightpermission[1]{}

\usepackage{tikz}
\usepackage{amsmath}
\usepackage[normalem]{ulem}
\usepackage{endnotes,microtype,xspace,fancyvrb,multirow}
\usepackage{xspace}
\usepackage{balance}
\usepackage{pifont}
\usepackage{anyfontsize}
\usepackage{booktabs}
\usepackage{fp}

\usepackage{listings}

\usepackage{balance}
\usepackage{textcomp}
\usepackage{stmaryrd}
\usepackage{enumitem}
\usepackage{array}
\usepackage{wrapfig}
\usepackage{makecell}

\usepackage{subcaption}
\usepackage{titlesec}

\usepackage{fontenc}
\usepackage{bbding}

\usepackage{xparse}

\newcommand{\sys}{HeteroK8s\xspace}
\newcommand{\os}{HeteroNet\xspace}
\newcommand{\xpod}{HeteroPod\xspace}
\newcommand{\faas}{\sys-FaaS\xspace}
\newcommand{\mesh}{\sys-Mesh\xspace}

\newcommand{\giptable}{\texttt{hetero-netns}\xspace}
\newcommand{\gsocket}{\texttt{hetero-socket}\xspace}

\newcommand{\codeword}[1]{\textcolor{black}{$\mathsf{#1}$}}

\newcommand{\myparagraph}[1]{\vspace {3pt}\noindent\textbf{\emph{#1}}}

\definecolor{hightcode}{rgb}{1.0,0.13,0.32}
\definecolor{myRubineRed}{RGB}{209, 0, 86}

\begin{document}


\title{HeteroPod: XPU-Accelerated Infrastructure Offloading for Commodity Cloud-Native Applications}

\author{Bicheng Yang}
\affiliation{%
  \institution{IPADS, Shanghai Jiao Tong University}
  \institution{Engineering Research Center for Domain-specific Operating Systems, Ministry of Education}
  \city{Shanghai}
  \country{China}
}

\author{Jingkai He}
\affiliation{%
  \institution{IPADS, Shanghai Jiao Tong University}
  \institution{Engineering Research Center for Domain-specific Operating Systems, Ministry of Education}
  \city{Shanghai}
  \country{China}
}

\author{Dong Du\textsuperscript{\Envelope}}
\thanks{\textsuperscript{\Envelope}Corresponding author: Dong Du (\url{Dd_nirvana@sjtu.edu.cn}).}
\affiliation{%
  \institution{IPADS, Shanghai Jiao Tong University}
  \institution{Engineering Research Center for Domain-specific Operating Systems, Ministry of Education}
  \city{Shanghai}
  \country{China}
}

\author{Yubin Xia}
\affiliation{%
  \institution{IPADS, Shanghai Jiao Tong University}
  \institution{Engineering Research Center for Domain-specific Operating Systems, Ministry of Education}
  \city{Shanghai}
  \country{China}
}

\author{Haibo Chen}
\affiliation{%
  \institution{IPADS, Shanghai Jiao Tong University}
  \institution{Engineering Research Center for Domain-specific Operating Systems, Ministry of Education}
  \institution{Key Laboratory of System Software (Chinese Academy of Science)}
  \city{Shanghai}
  \country{China}
}

\settopmatter{printacmref=false}
\settopmatter{printfolios=true}

\begin{abstract}

Cloud-native systems increasingly rely on infrastructure services (e.g., service meshes, monitoring agents), which compete for resources with user applications, degrading performance and scalability.
We propose \xpod, a new abstraction that offloads these services to Data Processing Units (DPUs) to enforce strict isolation while reducing host resource contention and operational costs.
To realize \xpod, we introduce \os, a cross-PU (XPU) network system featuring: (1) \emph{split network namespace}, a unified network abstraction for processes spanning CPU and DPU,
and (2) \emph{elastic and efficient XPU networking}, a communication mechanism achieving shared-memory performance without pinned resource overhead and polling costs.
By leveraging \os and the compositional nature of cloud-native workloads, \xpod can optimally offload infrastructure containers to DPUs.
We implement \os based on Linux, and implement a cloud-native system called \sys based on Kubernetes.
We evaluate the systems using NVIDIA Bluefield-2 DPUs and CXL-based DPUs (simulated with real CXL memory devices).
The results show that \sys effectively supports complex (unmodified) commodity cloud-native applications (up to 1 million LoC) and provides up to 31.9x better latency and 64x less resource consumption (compared with kernel-bypass design), 60\% better end-to-end latency, and 55\% higher scalability compared with SOTA systems.

\end{abstract}

\maketitle

\section{Introduction}
\label{s:intro}

Cloud-native computing has emerged as a prominent paradigm in cloud and datacenter environments,
utilizing \emph{containerization} to distribute and deploy applications~\cite{cncf, AWSLambda, IBMcloudFn, MSFunc, GoogleFunc}.
The fundamental unit of the execution environment, such as a \emph{Pod} in Kubernetes\footnote{In this paper, we consider the Pod as a generalized representation of the fundamental computational unit in cloud-native platforms.},
typically comprises one or multiple containers,
each containing a standalone image but sharing a network environment.
This approach encapsulates an application in a portable and self-contained way that can be deployed and distributed across diverse nodes.

\begin{figure}[t]
  \centering
 \setlength{\belowcaptionskip}{-10pt}
 \setlength{\abovecaptionskip}{0pt}
  \includegraphics[width=0.42\textwidth]{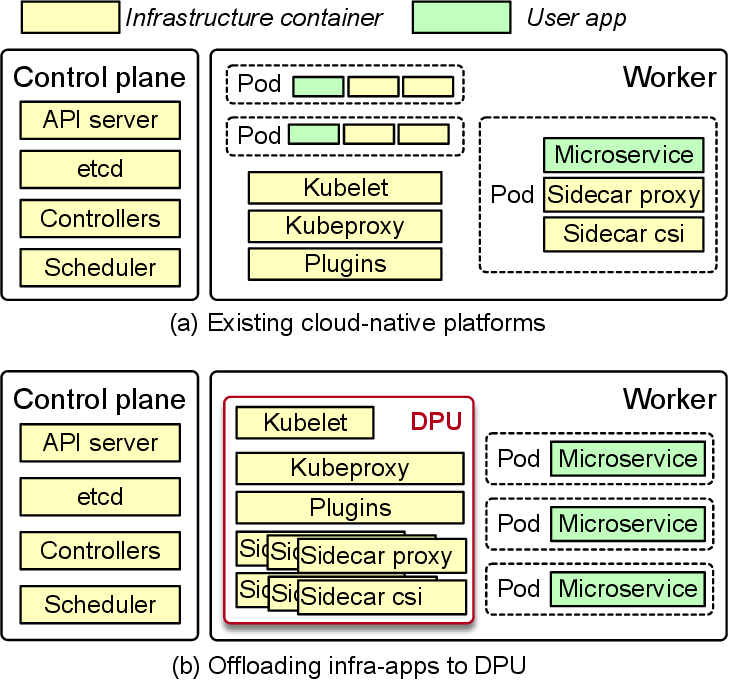}
  \caption{\textbf{Cloud-native platform with infrastructure containers and user apps.}}
  \label{fig:motiv-cloud-native-platform}
\end{figure}

One significant advantage of cloud-native apps compared to non-cloud-native apps is that they are developed with the support of extensive cloud infrastructure services,
including storage, network, security, and more.
These infrastructures are also deployed as containers.
There are two categories of infrastructure containers: global infrastructure containers and per-Pod infrastructure containers, as shown in Figure~\ref{fig:motiv-cloud-native-platform}-a.
For instance, the platform scheduler serves as one of the global infrastructure containers responsible for orchestrating requests to different Pods, while a sidecar proxy (e.g., Envoy~\cite{envoy}) represents a per-Pod infrastructure container present in each Pod to facilitate the networking functionalities of the application.

While infrastructure containers play a crucial role in delivering advantages in cloud-native computing,
they also introduce a non-trivial burden on resources, referred to as \emph{infra-burden} in the paper.
For example, service mesh~\cite{service-mesh} has emerged as a prominent pattern in cloud-native platforms for managing microservices,
employing a dedicated proxy (known as a \emph{sidecar-proxy}) to handle network traffic on behalf of the applications.
Since each microservice Pod contains a sidecar-proxy as an infrastructure container, these proxy containers can consume up to 25\% of the CPU and 33\% of memory resources on a node~\cite{10.1145/3620678.3624785}.
The infra-burden challenge has persisted for a considerable time~\cite{10.1145/3477132.3483565} and presents substantial difficulty in mitigation~\cite{10.1145/3694715.3695967, 10.1145/3694715.3695947, 10.1145/3694715.3695966}.

A recent noteworthy trend that could potentially offer relief is that \emph{cloud vendors are increasingly deploying more heterogeneous devices} to enhance scalability, performance, and cost-efficiency~\cite{amazon-f1, silberstein2013gpufs, kim2014gpunet, firestone2018azure, china-mobile-smartnic, huawei-smartnic, xilinx-fpga-huawei, ali-fpga, baidu-fpga}.
The Data Processing Unit (DPU) represents a novel category of heterogeneous devices that deliver scalable and energy-efficient computation capabilities~\cite{234944}, leveraging embedded processing units and an efficient NIC.
DPUs have the potential to significantly enhance instance density in a node (e.g., 50\% improvement in serverless computing~\cite{10.1145/3503222.3507732}).
These devices also offer enhanced security features through isolated memory and cores, along with boosted packet processing facilitated by numerous on-path hardware threads, such as the 256 hardware threads supported in Bluefield-3~\cite{bf-3}.
The expectation is that heterogeneous processing units like DPUs will be interconnected with rapid interconnect technologies like PCIe 6.0 and CXL~\cite{intel-cxl}, providing further performance enhancements.
Leveraging DPUs could potentially alleviate the infra-burden challenge.

\myparagraph{Goal: Offloading cloud-native infrastructure services to DPUs.}
In contrast to the conventional approach of colocating infrastructure containers and user apps on both the CPU and DPU~\cite{redhat-openshift},
we propose a promising (albeit challenging) architecture that involves decoupling these two types of cloud-native containers onto separate Processing Units (PUs), as illustrated in Fig.\ref{fig:motiv-cloud-native-platform}-b.
This strategy can effectively alleviate the infra-burden by preventing interference between infrastructure containers and app containers.
Nevertheless, separating infrastructure containers from app containers poses significant challenges.

\myparagraph{Challenge-1: Infrastructure and app containers in the same Pod are coupled with the same network namespace, dependent on the single-PU OS.}
Containers within a shared Pod utilize a will network namespace provided by the underlying OS kernel.
Consequently, processes in different containers share common network devices and can communicate across containers using ``localhost''.
However, offloading infrastructure containers to DPUs disrupts this shared network namespace functionality.
Sustaining local network abstraction while separating infrastructure and app containers onto different PUs poses a notable challenge.

\myparagraph{Challenge-2: Separating containers onto different PUs has a trade-off between communication performance and resource utilization.}
Infrastructure and app containers will communicate through network.
The separation may significantly increase communication latency (by 10x to 100x) compared to a single-PU design with Linux network stack as network packets should go through two NICs.
Advanced network solutions such as kernel-bypassed mechanisms~\cite{10.5555/2616448.2616493, rsocket, f-stack, 10.1145/3341302.3342071, libvma} like rsocket~\cite{rsocket} can provide socket-level APIs for compatibility and optimal performance to directly leverage hardware features like RDMA.
However, these solutions often encounter challenges related to suboptimal resource utilization~\cite{10.1145/3477132.3483569}, including memory wastage from pinned memory allocation for user-space links and CPU cycle wastage from user-mode polling.
Kernel-assisted approaches like SMC-R~\cite{rfc7609} have performance drawbacks due to context switching and data copying overheads.
Consequently, a cross-PU and network-based communication method that can deliver superior performance and maximize resource utilization is essential but currently unavailable.

\myparagraph{Challenge-3: Not all infrastructure containers are suitable for offloading to DPUs.}
The complexity of this challenge arises from the inherent nature of DPUs, which typically comprise a wimpy SoC (only) optimized for specific domain-centric tasks, such as network packet processing.
A simple design for offloading containers to DPUs may carry significant performance overhead, as suggested by prior works~\cite{wei2022comprehensive, 10.1145/3477132.3483565, 10.1145/3503222.3507732}.
As a result, system designers still lack effective practices to offload cloud-native infrastructure.

This paper introduces \xpod, a cross-PU Pod design that addresses the first two challenges through \os, which achieves transparent, high-performant, and resource-efficient XPU networking.
\os introduces two key techniques, namely \emph{split network namespace} (\giptable) and \emph{elastic and efficient XPU networking} (\gsocket).
\giptable provides a unified network abstraction for cross-PU containers.
The novelty of \giptable lies in its approach to \emph{re-do} network rules to establish a split network namespace, enabling direct or indirect communication as usual.
Second, \gsocket is a network stack design that attains kernel-bypassed performance, high resource efficiency, and socket compatibility.
\gsocket incorporates \emph{speculative allocation} to mitigate memory wastage, adopting a two-stage on-demand buffer allocation design that enables processes to speculatively allocate new buffers from a shared pool in a kernel-bypassed manner, with the kernel subsequently committing the allocation.
Additionally, \gsocket supports user-mode NAPI and facilitates cross-PU OS cooperation, leveraging page tables and exceptions to enable interrupts for U-mode libraries and diminish polling costs.

We utilize \xpod to implement \sys, a cloud-native system based on Kubernetes that supports serverless computing (\faas) and microservices with service mesh (\mesh~\cite{istio}).
\sys can effectively offload per-Pod infrastructure containers like sidecar proxies and global infrastructure containers like schedulers, while remaining general enough to offload suitable user apps in serverless to improve system density.
\sys acts as the guidelines to offload infrastructure containers for the third challenge.

We evaluate all the prototypes on real platforms with Nvidia Bluefield-2 DPUs and CXL-DPU simulators (based on our real CXL memory devices).
Evaluation results demonstrate that \gsocket can achieve up to 31.9x better latency and 64x less resource consumption.
Besides, \mesh can achieve 55\% higher scalability and 60\% better end-to-end latency compared to commercial Istio,
while \faas can achieve 20x lower communication latency compared to the baseline system and 4.45x less latency.
\sys can also offload control plane schedulers to DPU to cache global image states and implement an image-aware scheduling that can reduce avg latency by 14.4x.

\section{Motivation}
\label{s:challenge}

\subsection{Cloud-Native Apps and the Issues}

Cloud-native applications are usually deployed in a \emph{Pod}, which is a multi-container abstraction in Kubernetes and other platforms.
We classify cloud-native \emph{containers} into two types: infrastructure container (infra-container) and user-container, as shown in Fig.\ref{fig:motiv-cloud-native-platform}-a.
A promising benefit of cloud-native apps over non-cloud-native apps is that, \emph{cloud-native apps are developed with the supports of huge cloud infrastructure services}, including storage, network, security, etc.
For example, Envoy~\cite{envoy} is usually deployed as a infra-container by cloud vendors for each microservice (user-container) to manage network.
We highlight two issues (\textbf{I}) in nowaday cloud-native systems with such architecture.

\myparagraph{I1: Infrastructure containers bring significant burden on both CPU and memory.}
First, infra-containers introduce non-trivial resource burden.
For example, service mesh~\cite{service-mesh} is a popular pattern to manage microservices with a dedicated proxy, \emph{sidecar-proxy}, to handle network traffic on the behalf of the apps (Fig.\ref{fig:motiv-cloud-native-platform}-a).
We use the emoji-voting~\cite{emojivote} microservice to evaluate the resource costs introduced by sidecar containers on Kubernetes (detailed settings in \textsection\ref{s:case-mesh}).
Results show that proxy containers can consume up to 25\% of the CPU and 33\% of memory resources of a node.
Furthermore, proxy containers can consume a lot of resources even when RPS is low (RPS=20), which exacerbates resource interference further in a node.

\myparagraph{I2: Communication latency becomes more important for modular cloud-native apps.}
Microservice and serverless both are fine-grained and modular apps --- an app can be divided into tens to thousands of microservices or functions.
They rely on communication mechanism to cooperate.
As each service only takes few miliseconds for execution, the communication latency is important for end-to-end performance.
In fact, infra-containers like sidecar proxy can even increase the latency --- the latency can increase by 9.5x (RPS=20)--49.8x (RPS=3K) with proxy (Envoy)~\cite{10.1145/3620678.3624785}.
To avoid the costs, a common practice nowaday is to deploy a function chain (or highly related microservices) in the same node, and rely on local communication methods, e.g., IPC or domain socket, to reduce the latency~\cite{jianightcore, 10.1145/3503222.3507732, 273833, akkus2018sand}.
This leads to an important requirement, \emph{increasing the density of user apps, the higer the better.}
However, because of the interference of infra-containers, it is hard.

\begin{figure}[t]
\setlength{\belowcaptionskip}{-10pt}
\setlength{\abovecaptionskip}{0pt}
  \centering 
  \includegraphics[width=0.48\textwidth]{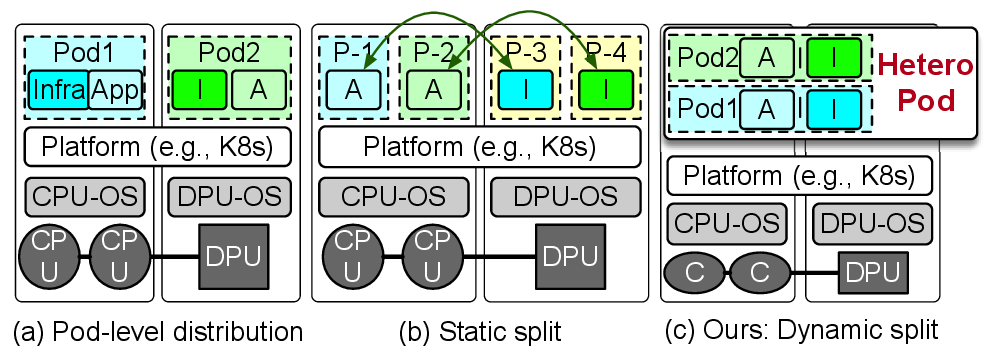}
  \footnotesize 
  \caption{\textbf{Cloud-native apps on CPU-DPU computers.}
  }
  \label{fig:intro-issue}
\end{figure}

\subsection{CPU-DPU Platforms for Cloud-Native Apps}
\label{s:dpu}

DPUs (data processing units)~\cite{netronomeagilio, mellanoxbf, marvel-octeon-sdk} are a new class of programmable devices that move and process data around the data centers.
It is an SoC and capable of running commercial OSes and performance-critical programs like network proxy.
DPU usually includes multi-core processing unit (e.g., 2.75GHz ARM cores in Bluefield-2 DPU) and efficient network interface.
Prior works have shown that DPU can improve the density of serverless functions~\cite{10.1145/3503222.3507732}, reduce the latency and resource costs for DFS~\cite{10.1145/3477132.3483565}, and enable energy-efficient microservices~\cite{234944}.

\myparagraph{Efforts using DPU for cloud-native apps.}
Commercial cloud platforms commonly use either 
\emph{pod-level distribution}, e.g., NVIDIA DPU Container~\cite{nvidia-doca-container} and Red Hat OpenShift~\cite{redhat-openshift},
or \emph{static split}~\cite{10.1145/3477132.3483565, li2020leapio, seshadri2014willow}, e.g., Azure~\cite{firestone2018azure},
as shown in Fig.\ref{fig:intro-issue}-a/b.
Pod-level distribution treats DPUs as individual servers and deploy unmodified apps directly on the DPUs.
However, apps may suffer significant performance degradation 
due to mismatches between their requirements and the hardware capabilities, e.g., a DPU usually has wimpy cores.
On the other hand, static split involves breaking down an app into fine-grained components that can be distributed across different PUs, leveraging communication for cooperation.
This approach can deliver great performance by deploying only suitable components on DPUs, such as computation-intensive components on CPU and network-intensive components on DPU.
However, this method typically requires an understanding of apps and hardware during development, making it challenging to adapt for generic apps and heterogeneous devices.
Despite these limitations, both methods are practical and can offer specific benefits in real-world industry apps.

Recent research efforts have led to the development of new heterogeneous frameworks for specific cloud-native scenarios, e.g., serverless computing~\cite{10.1145/3503222.3507732} and microservices~\cite{234944}.
However, these frameworks usually require app modifications, which limit their ability to support only simple apps (less than 1.8K lines of Node.js code in Molecule~\cite{10.1145/3503222.3507732}) and are \emph{not practical enough} for complex commodity applications like Envoy (a widely used service mesh proxy with 764K lines of C++ code).
This is a significant limitation given that many commodity cloud-native apps are distributed in binary or container image format, making it difficult to modify them.

\myparagraph{Our approach: dynamic split with \xpod.}
In contrast to prior methods, 
we propose \emph{dynamic cross-PU decoupling} --- transparently partitioning containers (even from the same Pod) across heterogeneous PUs during runtime,
as illustrated in Fig.~\ref{fig:intro-issue}-c. This architecture, termed \xpod (Cross-PU Pod), addresses critical limitations of conventional deployments.
First, isolating infrastructure containers on DPUs eliminates resource contention with application containers.
Second, DPU can help improve the density of a single node by also offloading suitable user-containers to DPU.
Last and also most important, unlike the static split approach, \xpod's dynamic split preserves existing Pod semantics, supporting unmodified commodity cloud-native apps.

\section{\xpod Design Overview}
\label{s:design-overview}

The core idea of \xpod is simple: selectively deploy infrastructure and app containers to DPUs, keeping others on CPUs. 
This container-granular deployment utilizes each container's self-contained nature to simplify decoupling. 
But the key challenge is \textbf{cross-container communication}, which depends on network interactions in cloud-native platforms.

Specifically, \xpod must address two technical challenges: 
(1) enabling transparent cross-PU networking to support unmodified applications (\textbf{Challenge-1} in \textsection\ref{s:intro}). (2) achieving high-efficiency cross-PU communication (\textbf{Challenge-2} in \textsection\ref{s:intro}).
To overcome these challenges, we design \os as the foundational technology for \xpod. \os introduces two key innovations for cross-PU (XPU) networking.

First, \os implements \giptable, a unified network abstraction for containers across PUs.
\giptable utilizes the global port space as an indicator to support cross-PU direct connection,
and leverages the observation that network filters/rules can be applied redundantly in both source and destination PUs (as long as they do not introduce new side effects) to enable cross-PU indirect communication.
As a result, \giptable creates a split network namespace, preserving communication semantics between components across PUs.
Second, \os introduces \gsocket, a kernel-library co-designed network stack that simultaneously achieves kernel-bypass performance, kernel-assisted resource efficiency, and socket API compatibility.
\gsocket employs two key mechanisms:
\emph{speculative allocation} is a two-stage buffer management strategy where processes temporarily claim buffers from shared pools (kernel-bypass manner), with the kernel committing allocations post-validation,
and \emph{user-mode NAPI} leverages cross-PU OS coordination to enable interrupt-driven processing in user-mode libraries, significantly reducing polling overhead.

\section{Design and Implementation of \os}
\label{s:design-hetero-net}

\begin{figure}[t]
  \centering
  \includegraphics[width=0.48\textwidth]{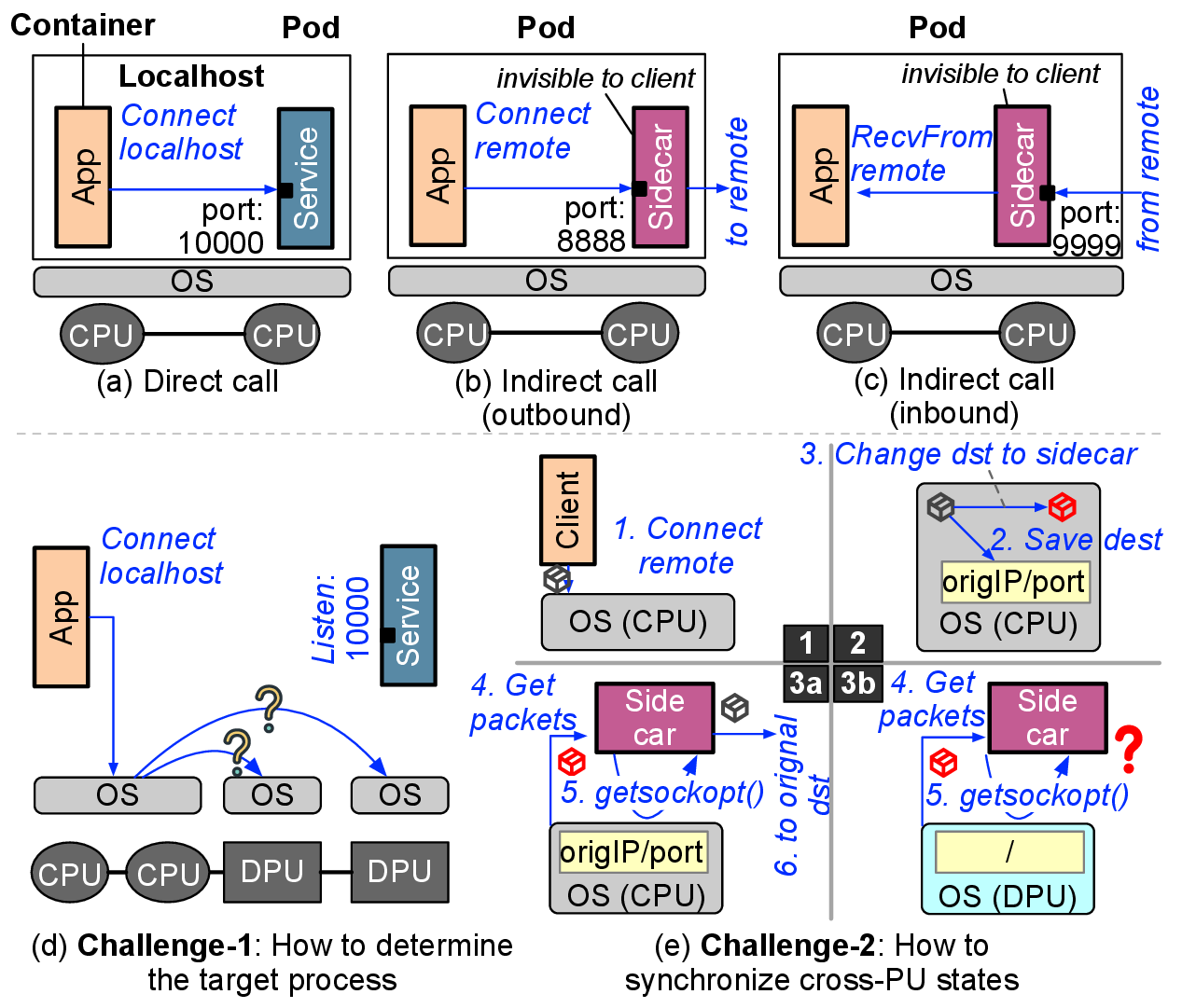}
 \setlength{\belowcaptionskip}{-15pt}
 \setlength{\abovecaptionskip}{0pt}
  \caption{\textbf{Two types of network calls and challenges of split network namespace}.
  }
  \label{fig:hetero-splitns}
\end{figure}

\subsection{Split Network Namespace (\giptable)}
\label{subs:hetero-iptable}

\myparagraph{Notation.}
Network namespaces in Linux~\cite{net-ns} provide isolation of network-related system resources,
including network devices, protocol stacks, IP routing tables, port numbers and so on.
We extend the concept to \emph{split network namespace}, which provides a unified and isolated network abstraction
in a CPU-DPU computer.
Split network namespace can be exclusively used by one container (multiple processes) or can be shared by multiple containers (e.g., Kubernetes Pod).

\myparagraph{Network calls.}
We can classify network requirements into two types (Fig.\ref{fig:hetero-splitns}-a/b/c): \emph{direct call} and \emph{indirect call}.
Direct call indicates two processes within the network namespace can communicate by using \codeword{localhost} and port number.
Indirect call refers to the scenario where a message is not directly transmitted to a local receiver, 
instead, it is initially intended for a remote receiver but is then forwarded to a local receiver due to packet filtering rules, such as iptables or routing.

\myparagraph{Challenges.}
First, challenges arise when multiple kernels manage processes within the ``same'' network namespace.
For direct call, the sender's OS kernel faces the difficulty of \emph{determining which kernel} is responsible for the receiver's processes.
Naively, the OS kernel on the sender side could forward packets to the kernel on the receiver side, but the issue lies in the lack of information regarding which processes are deployed on which PUs, as shown in Fig.\ref{fig:hetero-splitns}-d.

Second,
indirect call further introduces a new challenge called \emph{state synchronization}, where OS kernels on different PUs may explicitly synchronize certain states to process the packets accurately.
For example (Fig.\ref{fig:hetero-splitns}-e),
a sidecar proxy container is deployed to intercept incoming and outgoing packets from application containers.
Typically, this is accomplished by adding iptables rules to the network namespace of the Pod, redirecting packets sent by the application to the port where the proxy accepts them.
These iptables rules modify the destination IP and port information to match that of the proxy.
The OS kernels store the original destination IP/port information for future reference (Fig.~\ref{fig:hetero-splitns}-e, step-2/3).
Upon receiving the packet, the proxy container applies filtering processes and utilizes the \codeword{getsockopt()} syscall to retrieve the original destination IP/port information (Fig.~\ref{fig:hetero-splitns}-e, step-4/5).
A naive solution would be to add a similar rule on the sender side (i.e., the application container) to modify the destination IP and port, changing them to match the desired receiver's PU.
However, this poses a challenge because the proxy container (located on another PU) cannot retrieve the original IP and port information.
The sender's OS kernel (CPU in the fig), rather than the receiver's (DPU in the fig), stores this crucial data.
Although it is possible to synchronize the data between two PUs, it is very hard in practice because of the complicated rules and states.

\subsubsection{Port-based Reverse Mapping}
We have observed that in a split network namespace, processes distributed across different PUs still \emph{share the same logical port space}.
Each port number can only be exclusively used by one process.
Building on this observation, we propose the design of \emph{port-based reverse mapping}.
This approach involves maintaining a reverse mapping table that keeps track of the port numbers and their corresponding PUs. The mapping is globally shared by all \os daemons on each PU.
When a container needs to use a port number, the \os daemon checks the table to ensure that the port number is available. If the port is already in use, the operation will fail.

To avoid the overhead of port synchronization and checking during the data plane, \os adopts a plane decoupling approach.
When there are changes in port information, the \os daemon generates and applies a set of single-PU packet filtering rules, primarily utilizing iptables and routing in the prototype.
These rules automatically detect the \emph{remote port number} and employ DNAT (destination NAT) and SNAT (source NAT) techniques to correctly forward the packet to the appropriate PUs.
As a result, both the sender and receiver processes remain unaware of these underlying changes and can continue communicating using \codeword{localhost} with (almost) zero cost as if nothing has changed.

\subsubsection{Partially Idempotent Packet Rules}
To overcome the second challenges, we propose \emph{partially idempotent packet rules}.
Specifically, instead of performing the packet rules only in the source PU and synchronize the states latter (which is hard),
we transfer the original rules into two sets of new rules that will be applied to source and destination PUs.
These rules may have \emph{redundant} operations to a packet, but not introduce new side effects (i.e., \emph{idempotent} in this case).
States can be synchronized automatically by re-doing operations in both PUs.

For example, in the case of Fig.\ref{fig:hetero-splitns}-e, instead of simply changing the dst to receiver's IP/port and saving the orig-dst in the sender's PU,
we split it into three rules (simplified):
(1, on sender's PU) routing the pkt from sender's PU to receiver's PU which will not change the pkt,
(2, on receiver's PU) routing the pkt to the receiver's container,
and (3, on receiver's PU) changing the dst and save the orig-dst.
These rules perform redundant operations --- (1/2) and (3) both moving the pkt to the receiver PU, however, it will not change the behavior but achieve states synchronized.
As indirect call is achieved through network rules, \giptable can detect and replace them with the new rules for interception about incoming and outgoing traffic.

\subsection{Elastic and Efficient XPU Networking}
\label{subs:hetero-socket}

\begin{figure}[t]
 \setlength{\belowcaptionskip}{-10pt}
 \setlength{\abovecaptionskip}{0pt}
  \centering
  \includegraphics[width=0.45\textwidth]{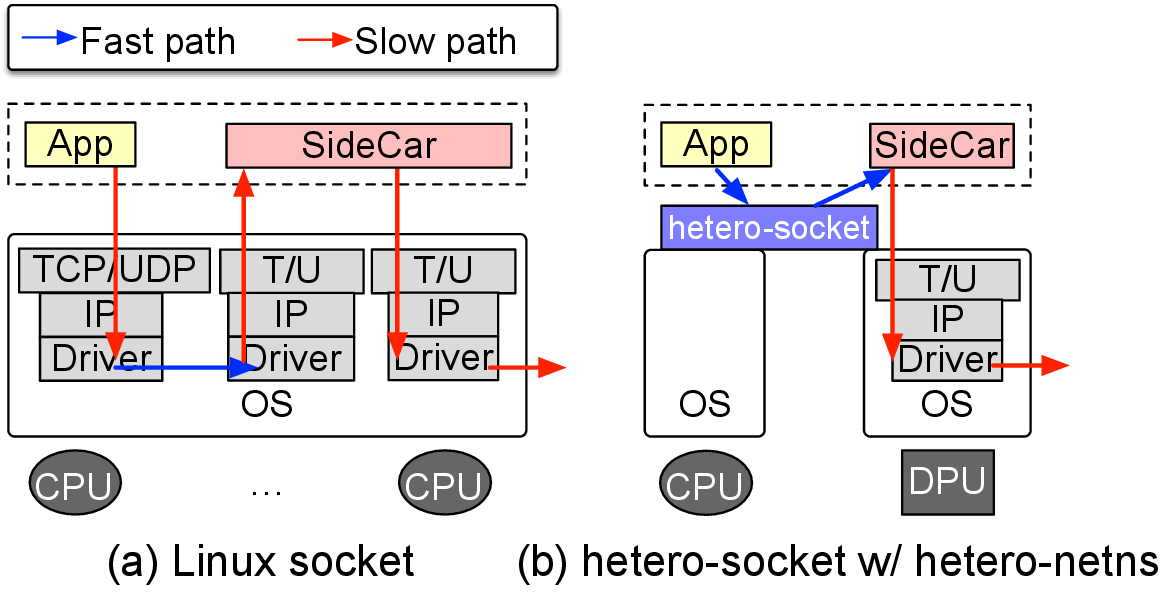}
  \caption{\textbf{Communication path with infra-containers.}
}  
  \label{fig:design-heterosocket-netstack}
\end{figure}

\subsubsection{Overview}

Traditional network stack introduces unncessary redundancy for XPU communication which causes longer latency and restricts the optimization opportunities of fast interconnect.
For example, in microservice with sidecar, a message has to go through the network stack three times to arrive at its destination,
as shown in Fig.~\ref{fig:design-heterosocket-netstack}.
Although there are kernel-bypass socket-compatible systems~\cite{f-stack, dpdk, libvma}, they suffer poor resource utilization because of pinned memory buffer and static allocation (\textsection\ref{subsubs:hetero-socket-challenges}).
\gsocket aims to achieve the same performance as prior kernel-bypass design, but overcoming the resource challenges in the XPU setting.

\myparagraph{Cross-PU interconnects as global shared memory.}
We summarize one key primitive for XPU communication, \emph{global shared memory} (or GShm),
which means there would be a shared memory that can be used as a channel for cross-PU communication.
This assumption already holds for existing DPUs~\cite{mellanoxbf, marvel-liquidio}, e.g., Bluefield-2 supports RDMA to allow CPU and DPU to share an explictly allocated region.
To be general to different interconnects, \gsocket introduces an IAL (interconnect abstraction layer), which abstracts the interconnect details but only expose GShm abstractions to the core of \gsocket.

\myparagraph{Decoupling of control and data paths.}
\gsocket relies on Linux network stack to establish a connection (with \giptable),
and will form a GShm-based channel for the best performance when available.
During runtime, the data-path APIs like \emph{read()} and \emph{write()} will be intercept using GShm when possible, but apps can still utilize control path operations like \emph{getsockopt()} to get the metadata for a connection.
When containers are deployed on different PUs, \gsocket relies on sockets and \giptable to establish a correct connection and then utilizes the GShm to boost communication.

\subsubsection{Challenges}
\label{subsubs:hetero-socket-challenges}
Although kernel-bypassed and shared memory-based design is quite common can achieve best performance by fully utilizing hardware capabilities,
\sys aims to further resolve two technical challenges which are not resolved yet.
First (\textbf{Pinned resource costs}), the allocation of shared memory regions is usually privileged operations that require the OSes of two PUs to negotiate using interconnect primitives.
This procedure is usually slow (e.g., RDMA~\cite{280746}), making dynamically extend the regions hard.
As a result, existing systems tend to \emph{reserve} a fixed size of region in OS for a connection to use, facing the trade-off between resource wastes and performance slowdown.
The issue is significant in real-world as there may be thousands of connections in a machine. 
Second (\textbf{Polling costs}), as OS kernel is bypassed, user-mode stacks are responsible to explictly poll the data, leading to an old trade-off between PU costs and long latency.

\subsubsection{User-mode Resource Management with Speculative Allocation}
\label{ss:spec-alloc}

\begin{figure}[t]
  \setlength{\belowcaptionskip}{-10pt}
  \setlength{\abovecaptionskip}{5pt}
  \centering
  \includegraphics[width=0.47\textwidth]{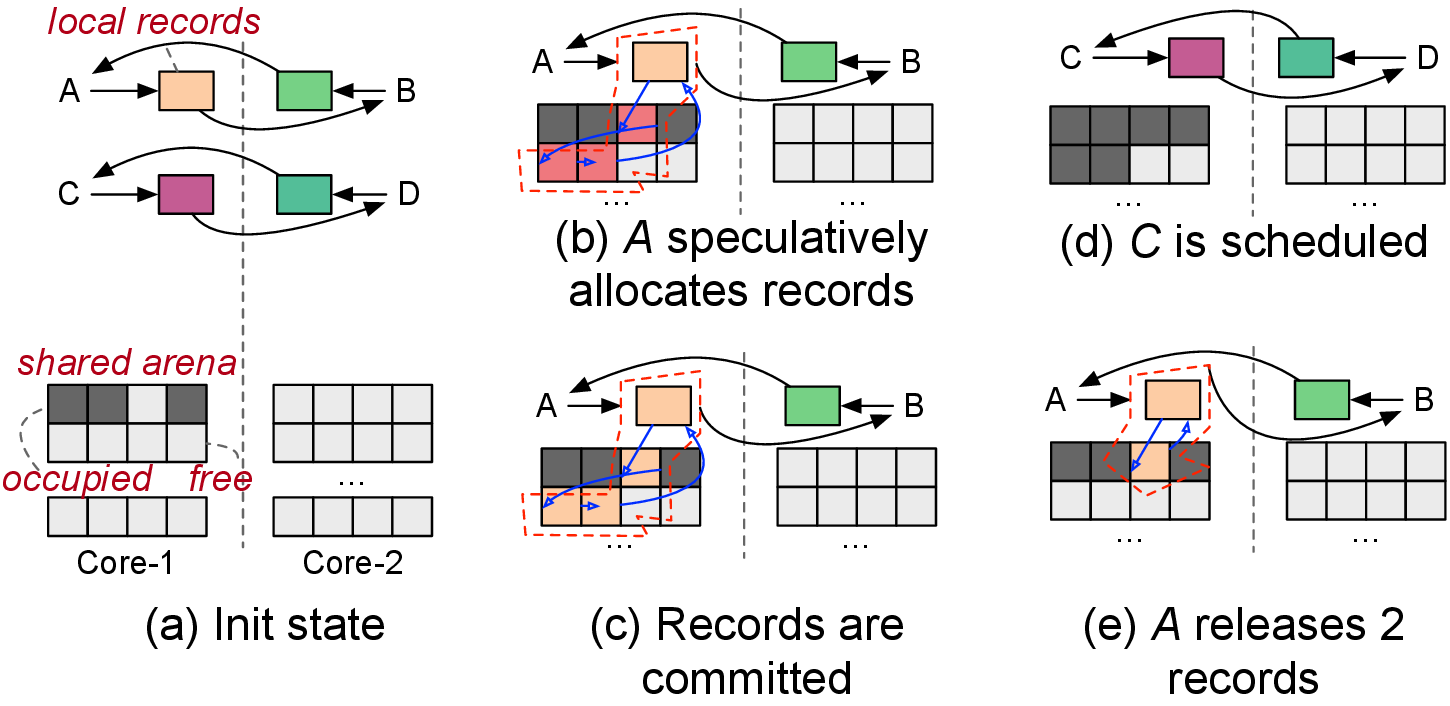}
  \caption{\textbf{Speculative allocation workflow} (a) record and arena, (b) speculative allocation, (c) kernel commit, (d) access control, (e) release.}
  \label{fig:design-heterosocket-umem}
\end{figure}

\gsocket employs an OS-library co-design for efficient memory management (Fig.~\ref{fig:design-heterosocket-umem}).
The core abstraction is the \emph{record} --- a page-aligned buffer with metadata (e.g., \emph{nextRecord}), unidirectionally accessible by either sender or receiver processes.
Each connection maintains two local records (one per direction) and leverages per-core \emph{shared arenas} containing dynamically allocatable records, managed as a list with \emph{headRecord} and \emph{tailRecord}.
This hybrid approach balances performance (local records) and scalability (shared arenas) while addressing three key challenges:

\myparagraph{Isolation \& synchronization.}
Shared arenas require strict access control: processes may only access their connection's records or their core's free records. \gsocket enforces this via memory mapping – during scheduling, the OS maps (1) local records, (2) occupied shared records for active connections, and (3) core-local free records. Per-core arenas eliminate synchronization overhead as free records are exclusively accessible during a process's time slice.

\myparagraph{Dynamic record chaining.}
Records form linked lists via a \emph{nextRecord} metadata. When a sender exhausts buffer space, it speculatively appends free records from its core's arena without kernel involvement (Fig.~\ref{fig:design-heterosocket-umem}-b).
The kernel later commits these allocations during context switches or exceptions (e.g., page faults) by remapping records as occupied (Fig.~\ref{fig:design-heterosocket-umem}c-d).
This two-phase approach --- user-space speculation followed by kernel ratification –-- achieves kernel-bypass performance with 64$\times$ less pinned memory (\S\ref{subs:eval-socket-microbench}).
For the receiver process of a connection, it may access a record that is newly allocated by the sender, which will trigger a page fault because of lacking permissions to access the free records.
In this case, the OS kernel will handle the page fault by mapping all newly allocated records to the receiver (also finishing the \textbf{commit}).
This trapping is rare in practice.

\myparagraph{Resource governance.} 
Rate-limiting policies prevent resource hoarding by bounding record usage per process.
The kernel enforces these during commit phases while preserving core-local allocation advantages.

\myparagraph{Design summary.} 
As a result, \gsocket can achieve similar resource management effectiveness as kernel-centric design except changing the gloabl resource pool to per-core pools (shared arena).
This change makes sense to avoid cross-core interferences and synchronization.
Compared with kernel-bypass network design with pre-reserved memory, \gsocket achieves similar performance as the trapping on the receiver side is very rare.
In fact, we even observe performance improvement as the dynamic resource of \gsocket can better dispatch resources among connections and better suit for varying loads.

\subsubsection{User-mode NAPI}
\label{ss:unapi}

\gsocket introduces the U-NAPI, a user-space adaptation of Linux's NAPI~\cite{linux-napi} that eliminates polling overhead through interrupt-driven event handling.
Traditional kernel-bypass designs waste CPU cycles polling for events --- U-NAPI instead dynamically switches between interrupt and polling modes based on traffic patterns, mirroring NAPI's adaptive behavior in OS kernels.

\myparagraph{Event management.}  
\gsocket keeps standard \codeword{epoll} semantics and adds two key syscalls: \codeword{enable\_notify\_gshm} and \codeword{disable\_notify\_gshm}. 
These allow applications to register interest in events (e.g., buffer availability) with the kernel.
\codeword{enable\_notify\_gshm} supports two modes (using different flags): \codeword{sync} and \codeword{async}.
The \codeword{sync} flag blocks threads until events occur (ideal for blocking I/O), while \codeword{async} triggers signal handlers that disable interrupts and enable polling for latency-sensitive phases.
The key innovation here is that U-NAPI leverages OS kernel to simulate the hardware to provide interrupt capability for U-mode event management.

\myparagraph{Integration with network operations.}  
In practice, U-NAPI optimizes three common scenarios.
First, a network \codeword{write} (or \codeword{send}) operation may block and turns to polling because the buffer of a connection is full.
In this case, the \gsocket uses \codeword{enable\_notify\_gshm(token, sync | W\_flag)},
which blocks in the kernel until the connection is ready for write.
Second, a similar case is blocking \codeword{read} (or \codeword{recv}) on an empty connection.
The \gsocket can use \codeword{sync} mode with \codeword{R\_flag}.
Last, to handle the case to add a new event to an epoll instance,
\gsocket can use \codeword{async} mode to monitor events, i.e., \codeword{enable\_notify\_gshm(token, async | events)}.
Then it can use the signal handler to handle any event updates, disabling notification, and turn to polling for performance.

\myparagraph{Cross-PU event propagation.}  
U-NAPI uses two cross-PU event delivery techniques.
First, \emph{proactive notification} has the remote side of a connection proactively tell the remote (or same PU) kernel of a new event. 
E.g., in blocking write, when the remote receiver reads data, checks status, and finds the sender blocked, it uses a syscall to tell the kernel the connection status changed, which eventually alerts the sender space may be available.
Second, \emph{Notify-on-RW (NoRW)} makes the kernel tell the remote (or same) kernel to revoke connection record permissions (e.g., no read-permission in the write case). This causes page faults on the receiver's read, which are sent to the sender's kernel to notify of the connection status change. 
Though cross-PU latency exceeds local cases, apps retain the flexibility to revert to polling for critical paths.

\subsection{Implementation}
\label{subs:imp}

\myparagraph{\giptable}.
\giptable runs as a daemon process on each PU (Linux) to monitor local networking rules and coordinate with other PUs within the same computer.
It maintains ports and intercepts network rules within a network namespace.
\giptable leverages existing networking techniques, e.g., nsenter, NAT, policy routing, and iptables.
In \sys, we extend Kubernetes' CNI with \giptable.

\myparagraph{\gsocket.}
\gsocket is written in C/C++.
We leverage \codeword{LD\_PRELOAD}~\cite{ld-preload} to replace socket-related functions (implemented in libc) of an application at runtime.
Besides, we implement global shared memory (GShm) based on CXL (on CXL-DPU) and RDMA (on Bluefield-2).
The \gsocket implements a full set of Linux socket interfaces. 
We extend Linux kernel to support the speculative allocation and the interfaces of U-NAPI.

\myparagraph{Multi-ISA supports.}
We use the multi-platform container image feature~\cite{multi-platform} to support multi-ISA PUs in the same computer, which is a standard feature in Docker/Kubernetes.

\myparagraph{Security.}
\os is implemented as a per-process library now, which will maintain the global shared memory between two entites in userspace.
As each process can only access the shared memory for its own connections, \os will not introduce new security flaws to attack other processes.
A malicious process may break the connection, but can not cause other issues like overflow because of careful pointer operations.
Moreover, \os allows administrator to configure which connections are allowed to utilize the \gsocket by a configuration file, 
which is similiar like firmware to explictly specify the allowed list.

\myparagraph{Experimental setup.}
We use two settings for evaluation.
First (\textbf{Bluefield-DPU}), we use two servers, each with 2xE5-2650 v5 CPU (12cores, 2.2GHz each CPU), 96GB DDR4-1600 memory, and one Nvidia Bluefield-2 DPU devices.
Each Bluefield-2 DPU has 8 ARM Cortex-A72 cores (2.75GHz), 16GB DDR-1600 DRAM and ConnectX-6 (2x 100Gbps RDMA ports) NIC.
We also validate the results on Bluefield-3.
Second (\textbf{CXL-DPU}), 
we use a dual-socket server with Intel Xeon Platinum 8468V (2.9 GHz, 48 cores) and a 32GB 4800MT/s DDR5 RDIMM is connected through a CXL memory expansion card (CXL 2.x).
We implement CXL-based DPU with two VMs (representing CPU and DPU) and utilize the real CXL device (configured as \codeword{dax} dev) as the GShm.

\begin{figure*}[t]
    \centering
   \setlength{\belowcaptionskip}{-5pt}
   \setlength{\abovecaptionskip}{0pt}
    \includegraphics[width=0.99\textwidth]{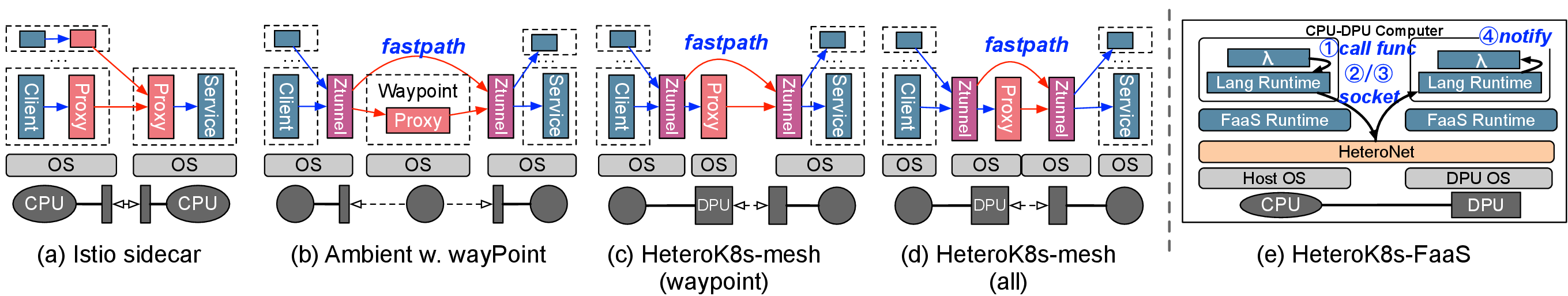}
    \caption{Service mesh with different architectures (a/b), \mesh (c/d) and \faas (e).}
    \label{fig:intro-wingman}
  \end{figure*}

  \begin{figure*}[htb]
   \setlength{\belowcaptionskip}{-5pt}
   \setlength{\abovecaptionskip}{0pt}
  \begin{minipage}[t]{0.49\linewidth}
    \centering 
    \includegraphics[width=0.99\textwidth]{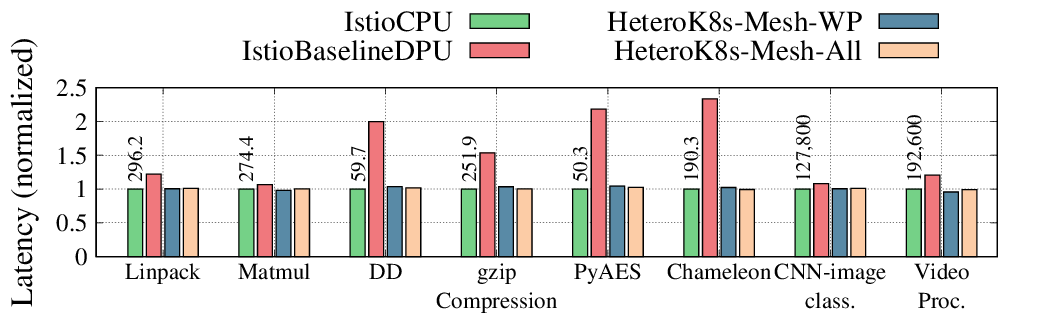}
    \footnotesize
    \textbf{(a) FuncBench.}
    \end{minipage}
  \begin{minipage}[t]{0.49\linewidth}
    \centering 
    \includegraphics[width=0.99\textwidth]{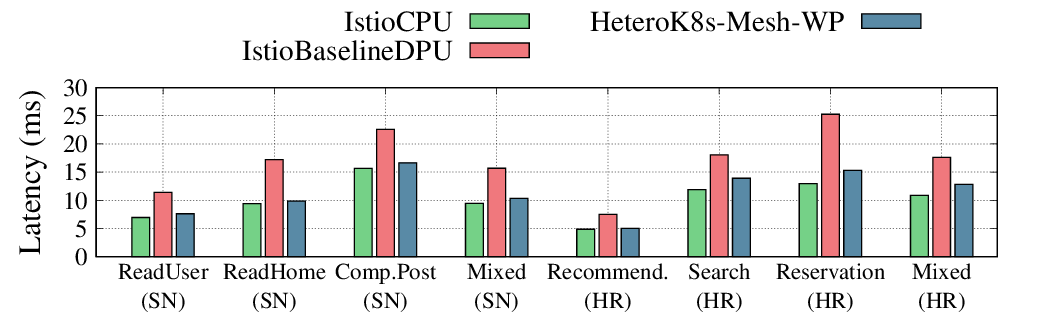}
    \footnotesize
    \textbf{(b) DeathStarBench.}
    \end{minipage}
  
   \begin{minipage}[t]{0.24\linewidth}
    \centering 
    \includegraphics[width=0.99\textwidth]{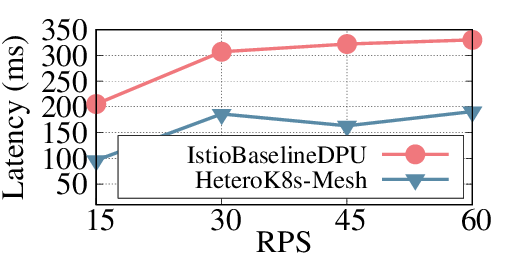}
    \footnotesize
    \textbf{(c) Avg latency.}
    \end{minipage}
    \begin{minipage}[t]{0.24\linewidth}
    \centering 
    \includegraphics[width=0.99\textwidth]{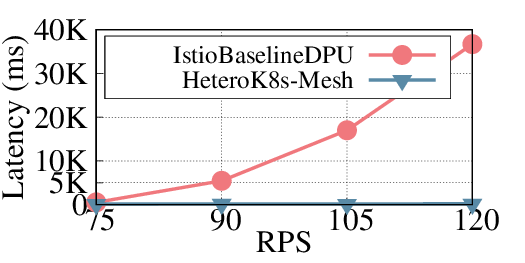}
    \footnotesize
    \textbf{(d) Avg latency (high RPS).}
    \end{minipage}
    \begin{minipage}[t]{0.24\linewidth}
    \centering 
    \includegraphics[width=0.99\textwidth]{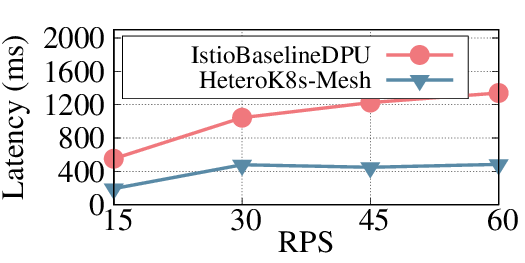}
    \footnotesize
    \textbf{(e) P99 latency.}
    \end{minipage}
    \begin{minipage}[t]{0.24\linewidth}
    \centering 
    \includegraphics[width=0.99\textwidth]{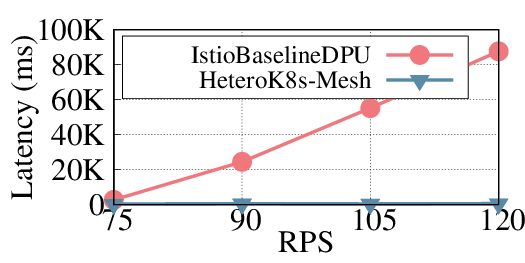}
    \footnotesize
    \textbf{(f) P99 latency (high RPS).}
    \end{minipage}
      \caption{\textbf{Microservice with service mesh (Bluefield2-DPU).} \textit{In deathstar, SN means Social Network, HR means Hotel Reservation. We present end-to-end latency. The unit is milliseconds unless explicitly declared.} }
    \label{fig:eval-mesh-big}
  \end{figure*}

\section{\sys: XPU Cloud-Native System}
\label{s:case}

We present \sys, a cloud-native system based on Kubernetes that utilize \os to effectively and efficiently utilize DPUs under different scenarios.
We present three case studies,
offloading worker node's infra-apps (\textsection\ref{s:case-mesh}),
offloading apps (\textsection\ref{s:case-FaaS}),
and offloading control plane's infra-apps (\textsection\ref{s:case-scheduler}).

\subsection{Case Study: Service Mesh and Microservice}
\label{s:case-mesh}

Service mesh is a representative pattern which injects a \emph{sidecar} container (proxy) into each microservice Pod, leading to both resource and communication latency challenges.
We illustrate how \sys can utilize \os to optimize the costs caused by proxy-related infra-containers.

\myparagraph{Baseline systems with Istio.}
Besides the basic proxy design (Fig.\ref{fig:intro-wingman}-a),
Istio~\cite{istio} (SOTA mesh system) supports an advanced architecture, \emph{waypoint} sidecar~\cite{istio-ambient} (Fig.\ref{fig:intro-wingman}-b), that can effectively mitigate costs.
Specifically, it decouples the sidecar into a \emph{zero-trusted tunnel} (ztunnel) and a \emph{waypoint sidecar}.
Ztunnel can be shared by all apps on a node for L4 policies, while a waypoint sidecar is dedicated to an app (for L7 policies) and can be scheduled to other nodes.
The decoupling can save resources on the apps' node, however, it also increases the latency for the normal path resulting from the scheduling of waypoint sidecars on a third server.

\myparagraph{\mesh.}
We design \mesh to balance performance and scalability with efficient offloading of sidecars to DPUs with \os.
\mesh supports two deployment methods:
(1) offloading all proxy-related infra-containers (ztunnel and waypoint) to DPU (Fig.\ref{fig:intro-wingman}-d), so that CPU can focus on running microservice apps,
and (2) offloading waypoint containers to DPU (Fig.\ref{fig:intro-wingman}-c).
It will launch new ztunnel and waypoint containers to CPU when DPU devices fail or their resources are insufficient to ensure the availability and QoS of microservices.
If the client and server microservices are on the same node, \mesh will deploy the ztunnel and waypoint containers in the CPU when possible.
\mesh will utilize \giptable to support indrect communication between apps and proxy, and rely on \gsocket to speed up the communication latency.

\myparagraph{Methodology.}
We deploy DeathStarBench~\cite{10.1145/3297858.3304013} and FunctionBench~\cite{kim2019practical} as microservices, and deploy each microservice with an Istio waypoint (using Envoy) as gateway to evaluate the performance of \mesh.
All incoming and outgoing traffic of microservices will be intercepted and processed by the ztunnel on the node.
We compare \mesh with two systems: IstioCPU and IstioBaselineDPU.
IstioCPU will only deploy microservice Pods on the CPU, which represents the best performance in many cases, but can not effectively utilize DPUs to improve the scalability.
IstioBaselineDPU is implemented based on Istio to deploy microservice Pods on both CPU and DPU (priorly scheduled to DPU).
It represents existing approaches utilizing DPU using pod granularity.

\myparagraph{End-to-end performance.}
We evaluate microservices' end-to-end latency under two baseline systems, \sys-WP (Fig.~\ref{fig:intro-wingman}-c) and \sys-All (Fig.~\ref{fig:intro-wingman}-d). 
To show \emph{dynamic split}'s benefits, \mesh does not use \gsocket here.

The evaluation results for FunctionBench are shown in Fig.\ref{fig:eval-mesh-big}-a.
Compared with the IstioBaselineDPU, \mesh can shorten microservices' end-to-end latency up to 60\% (dd, pyaes, chameleon). 
The performance of all apps is improved after deploying with \mesh.
Compared to only using CPU, using \mesh-WP to offload waypoint to DPU only brings end-to-end latency overhead of less than 5\%.
Using \mesh-All to offload ztunnel and waypoint to DPU can saves more CPU resources, and only brings end-to-end overhead of 1\%-2\%(dd, pyaes)  and $<$1\% for all other apps.
The reason why the end-to-end latency overhead of using \mesh-All is smaller than that of \mesh-WP is that after the service mesh infrastructure is fully offloaded to DPU, the microservices running on the CPU are subject to less competition for CPU and microarchitecture resources (such as cache and TLB), which helps to shorten the running time of microservices.
The evaluation results for DeathStar are shown in Fig.~\ref{fig:eval-mesh-big}-b.
We do not offload ztunnel here as microservices in DeathStar have short runtime latency.
\mesh can achieve 29\%--74\% better latency compared with BaselineDPU, and it incurs only $<$1.2ms latency costs (on average) compared with CPU (the ideal performance without offloading).

\myparagraph{Scalability.}
We compare the scalability under high density.
Specifically, both \mesh and IstioBaselineDPU will use all DPU and most CPU resources to deploy Pods.
Each microservice pod is paired with a waypoint proxy.
The total number of microservices and waypoints is the same (70 instances), and the same resources (CPU and memory) are allocated, i.e., 128MB memory and 100 mCPU per waypoint, and 256MB memory and 300 mCPU per app.
We use Pyaes as the testcase.
Requests are evenly load-balanced to all microservices in a random manner using Nginx.

The average latency results are presented in Fig.\ref{fig:eval-mesh-big}-c/d and the P99 latency in Fig.\ref{fig:eval-mesh-big}-e/f.
Under different RPS, microservices deployed using \mesh achieve better performance than the IstioBaselineDPU --- 39\% improved end-to-end latency.
When the RPS rises, the latency of microservices deployed using the baseline scheme increases rapidly,
while the latency using the \mesh scheme can remain relatively stable.
This means that \mesh has better scalability than baseline.

\begin{figure}[tb]
\begin{minipage}[t]{0.49\linewidth}
  \centering 
  \includegraphics[width=0.99\textwidth]{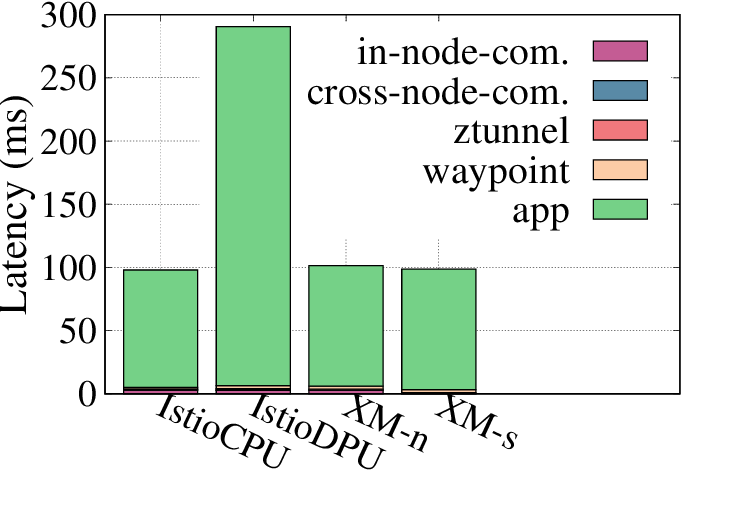}
  \footnotesize \\[-10pt]
  \textbf{(a) Large application.}
  \end{minipage}
\begin{minipage}[t]{0.49\linewidth}
  \centering 
  \includegraphics[width=0.99\textwidth]{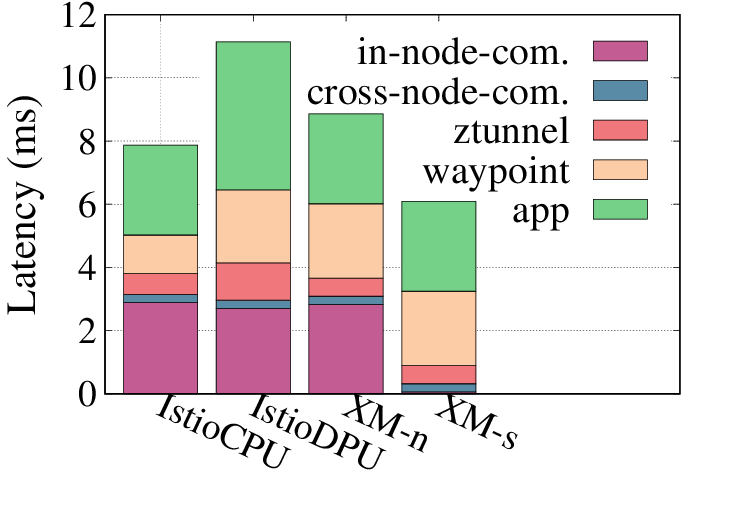}
  \footnotesize \\[-10pt]
  \textbf{(b) Small application.}
  \end{minipage}
 \setlength{\belowcaptionskip}{-15pt}
 \setlength{\abovecaptionskip}{2pt}
  \caption{\textbf{Performance Breakdown (Bluefield2-DPU)}. \textit{XM-n: \sys-Mesh (hetero-netns).} \textit{XM-s: \sys-Mesh (hetero-sock).}}
  \label{fig:eval-mesh-breakdown}
\end{figure}

\myparagraph{Performance breakdown.}
\label{subs:eval-scale}
We present a breakdown using large and small apps.
We break the end-to-end latency into five parts: in-node (including inter-PU and cross-PU) communication latency, cross-node communication latency, and processing latency of ztunnel, waypoint and application.
As shown in Fig.\ref{fig:eval-mesh-breakdown},
for large app, \mesh (\giptable) can achieve great performance --- 2x better than IstioDPU and almost the same as the IstioCPU performance.
\gsocket brings little improvement here because the processing costs dominate the latency here.
For small app, \mesh (\giptable) outperforms IstioDPU. But it has 16\% higher latency than IstioCPU due to the increased latency on waypoint.
\mesh (+\gsocket) can further improves the latency by optimizing the communication latency, and achieve even better performance compared with IstioCPU.

\subsection{Case Study: Serverless Computing (FaaS)}
\label{s:case-FaaS}

Recent works~\cite{10.1145/3472883.3486972, 10.1145/3503222.3507732, copik2022rfaas, wei2022provisioned} have tried to explore heterogeneous devices (e.g., DPU) for serverless to achieve better scalability and performance, e.g., Molecule~\cite{10.1145/3503222.3507732}. 
However, they usually require non-trivial modifications to both the serverless runtime and applications.

\myparagraph{\sys-FaaS.}
\sys supports serverless computing on CPU-DPU computers without modifying binaries of functions, language runtimes, and serverless runtime (Fig.\ref{fig:intro-wingman}-e).
Unlike Molecule which utilizes one serverless runtime to manage both CPU and DPU, \faas takes different PUs as standalone devices, but incorporating the \os to achieve efficient communication between functions transparently.
As Molecule's function are single-container sandboxes, \faas does not apply the \giptable.

We implement \faas based on the Molecule's open-sourced homogeneous implementation, and utilizes its modified runc which support container fork for low latency startup.
\faas will preload \os for each function instance, but configure a \emph{configuration file} to only enable \gsocket between instances in the same function chain, i.e., one-function applications will not utilize \os.
\faas assumes the global gateway will schedule a function chain in a one heterogeneous computers.
We compare \faas with Molecule and Molecule-homo (a homogeneous version for CPU-only computer).

\begin{figure}[t]
 \setlength{\belowcaptionskip}{-10pt}
 \setlength{\abovecaptionskip}{0pt}
  \begin{minipage}{0.49\linewidth}
    \centering 
    \includegraphics[width=1\linewidth]{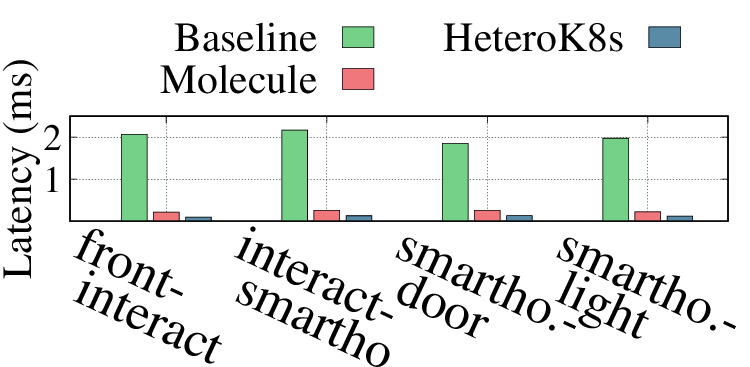}
    \footnotesize
    \textbf{(a) CPU to CPU}
  \end{minipage}
  \begin{minipage}{0.49\linewidth}
    \centering 
    \includegraphics[width=1\linewidth]{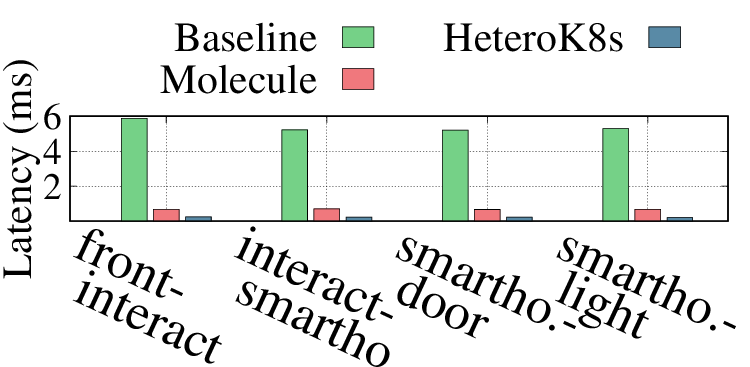}
    \footnotesize
    \textbf{(b) DPU to DPU}
  \end{minipage}

  \begin{minipage}{0.49\linewidth}
    \centering 
    \includegraphics[width=1\linewidth]{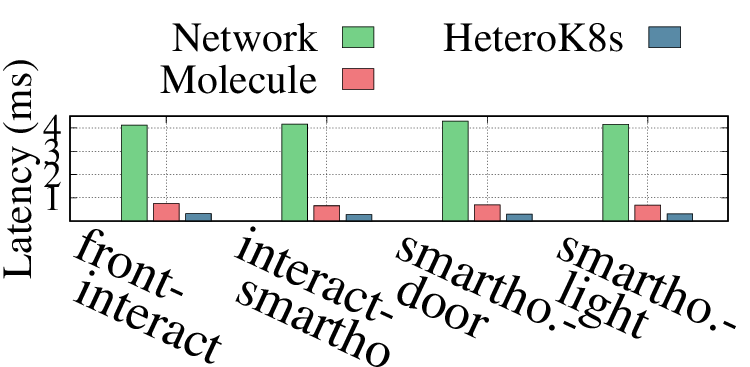}
    \footnotesize
    \textbf{(c) CPU to DPU}
  \end{minipage}
  \begin{minipage}{0.49\linewidth}
    \centering 
    \includegraphics[width=1\linewidth]{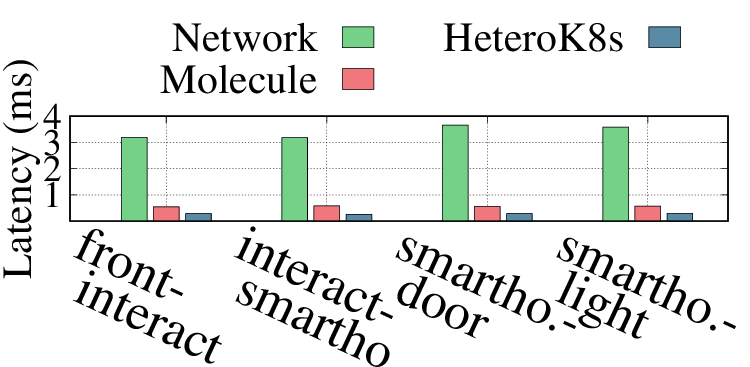}
    \footnotesize
    \textbf{(d) DPU to CPU}
  \end{minipage}
  \caption{\textbf{Serverless DAG communication latency.} \textit{\faas achieves 2x lower latency compared with state-of-the-art system, Molecule.}}
  \label{fig:eval-runtime-micro-dag}
\end{figure}

\myparagraph{Microbenchmarks.}
We perform microbenchmarks to analyze the communication latency between two functions in a chain.
We take Alexa skills~\cite{socc20-serverlessbench} as an example,
and compare \faas with Molecule (using nIPC) and the baseline (Molecule's homogeneous version using Node.js Express).
We consider four cases, including CPU-only, DPU-only, CPU-to-DPU and DPU-to-CPU.
The results are shown in Fig.\ref{fig:eval-runtime-micro-dag}.
Molecule already achieves about 10x lower latency compared with the Baseline because of its efficient nIPC-based communication.
\faas can achieve even better performance (about 2x) because Molecule requires one additional IPC call to invoke the Hetero-Shim.
The results of Molecule is better than the reported data~\cite{10.1145/3503222.3507732} because of the better hardware settings in our environment.

\begin{figure}[t]
\setlength{\belowcaptionskip}{-10pt}
\setlength{\abovecaptionskip}{0pt}
\begin{minipage}[t]{0.66\linewidth}
  \centering 
  \includegraphics[width=1\linewidth]{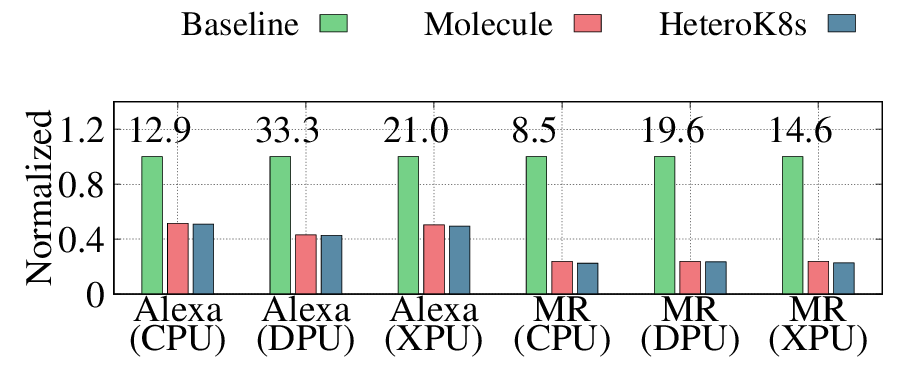}
  \footnotesize    
\textbf{(a) Apps on Bluefield-2}
\end{minipage}  
\begin{minipage}[t]{0.33\linewidth}
  \centering 
  \includegraphics[width=1\linewidth]{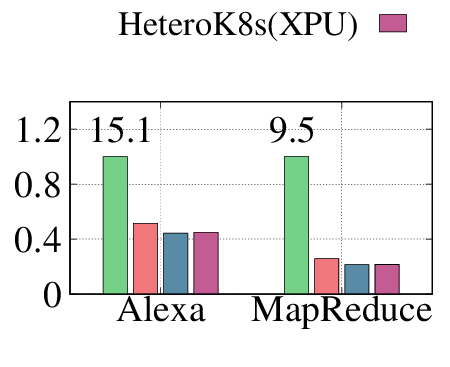}
  \footnotesize    
\textbf{(b) Apps on CXL-DPU}
\end{minipage}  
\caption{\textbf{Chained Applications.}
	\textit{MR for MapReduce. XPU for Cross-PU.}
	}
  \label{fig:eval-faas-dag-apps}
\end{figure}

\myparagraph{Applications.}
We analyze end-to-end latency of two chained applications from ServerlessBench and FunctionBench, Alexa and MapReduce.
Alexa comprises five Node.js functions and MapReduce includes three Python functions.
We compare three systems, Baseline (i.e., Molecule-homo), Molecule, and \faas under three cases, CPU, DPU and cross-PU.
In the case of cross-PU, we ensure that all inter-function calls are cross PU, e.g., in Alexa, the 1st, 3rd, and 5th functions are in the host CPU, and the 2nd and 4th functions are in the DPU.
All the instances are warm-boot (without cold startup costs).
The results are shown in Fig.\ref{fig:eval-faas-dag-apps}.
Molecule can achieve 1.94--2.32x less end-to-end latency in Alexa and 4.20--4.21x less latency in MapReduce,
while \faas can achieve better latency, achieving 1.96--2.34x less end-to-end latency in Alexa and 4.19--4.45x less latency in MapReduce compared with the baseline.

\begin{figure}[t]
\setlength{\belowcaptionskip}{-10pt}
\setlength{\abovecaptionskip}{0pt}
\begin{minipage}[t]{0.49\linewidth}
  \centering 
  \includegraphics[width=1\linewidth]{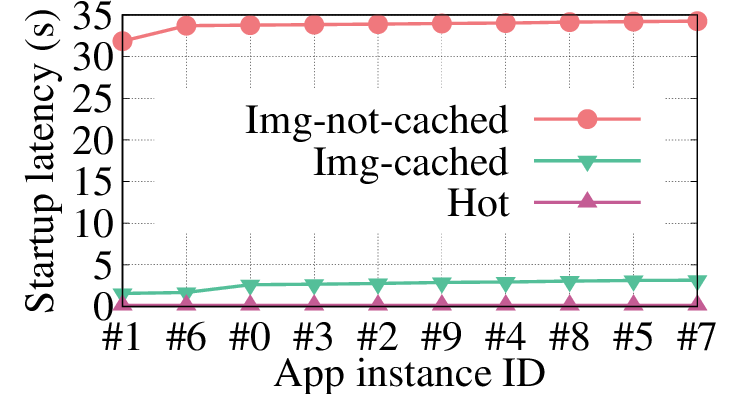}
  \footnotesize    
\textbf{(a) Motivation (Knative)}
\end{minipage}  
\begin{minipage}[t]{0.49\linewidth}
  \centering 
  \includegraphics[width=1\linewidth]{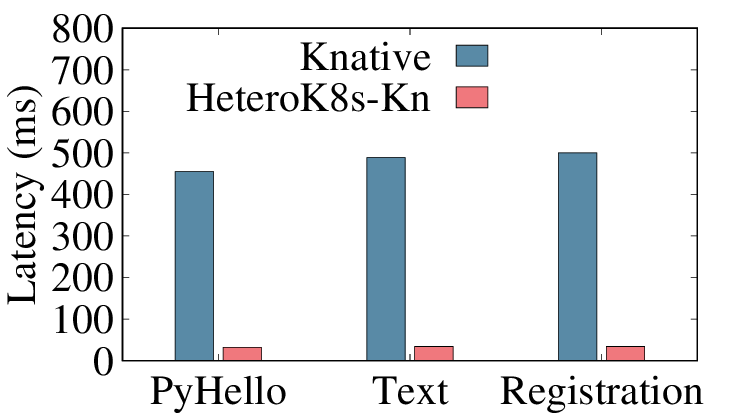}
  \footnotesize    
\textbf{(b) Optimization results}
\end{minipage}  
\caption{\textbf{Offloading scheduler for image-aware scheduling.}
  }
  \label{fig:eval-scheduler}
\end{figure}

\subsection{Case Study: Offloading Scheduler}
\label{s:case-scheduler}

\myparagraph{Motivation.}
Startup latency is one of the most important metrics for cloud-native apps~\cite{socc20-serverlessbench}.
Although there are many efforts to utilize C/R, fork, and caching, to optimize the container startup latency, we observe significant challenges from \emph{image pull},
i.e., if a request is scheduled to a node without the prepared image, the platform will pull the image first and then launch the instance.
To analyze the costs,
we evalaute the response latency in Knative~\cite{knative}, a widely-used FaaS system, with Kperf.
As shown in Fig.\ref{fig:eval-scheduler}-a, the latency of image not-cached is about 10x longer than the image-cached case.

\myparagraph{Idea and challenges.}
Cloud-native platforms usually rely on a global scheduler to schedule a Node for a request, e.g., kube-scheduler in K8s.
An idea is \emph{image-aware scheduler}, which prefers to assign a request to a Node with prepared images,
e.g., K8s supports image-locality policy~\cite{image-locality}.
However, this requires scheduler to maintain the image information of all nodes.
For a real-world environment, caching all image information requires 1GB--10GB resource.
Besides, updates of image status will notify the scheduler,
making it the bottleneck component in the control plane.

\myparagraph{Solution.}
\sys can offload the image cache in the scheduler to the DPU, utilizing the redundant resources to maintain images.
Moreover, as DPU is more closer to the network, it is more reasonable to use DPU to update image status.
Based on the offloading, we design an image and tempalte (for container fork~\cite{10.1145/3503222.3507732}) aware policy (as a kube-scheduler plugin), which will prioritize nodes with template instances and cached images.
We support Knative (w./o. modification) on \sys with the offloaded scheduler and new policy.
As shown in Fig.\ref{fig:eval-scheduler}-b, we analyze the avg latency in the whole cluster (four machines), and \sys-Kn can reduce the avg latency from 481ms to 33ms (14.4x).

\section{Evaluation of \os}
\label{s:eval}

\begin{figure}[tb]
 \setlength{\belowcaptionskip}{-10pt}
 \setlength{\abovecaptionskip}{0pt}
\begin{minipage}[t]{0.48\linewidth}
  \centering 
  \includegraphics[width=0.99\textwidth]{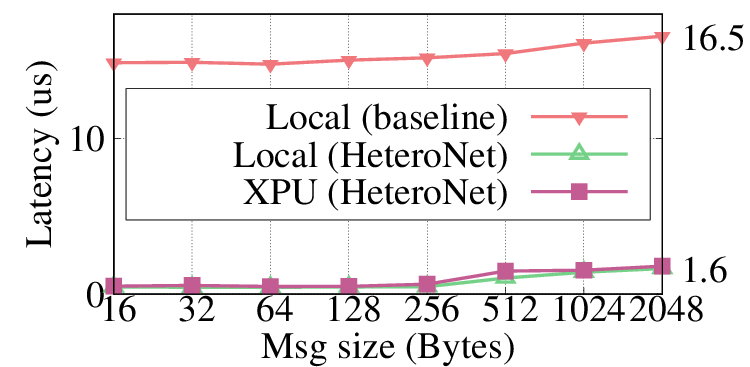}
  \footnotesize 
  \textbf{(a) Latency (small msg).}
  \end{minipage}
\begin{minipage}[t]{0.48\linewidth}
  \centering 
  \includegraphics[width=0.99\textwidth]{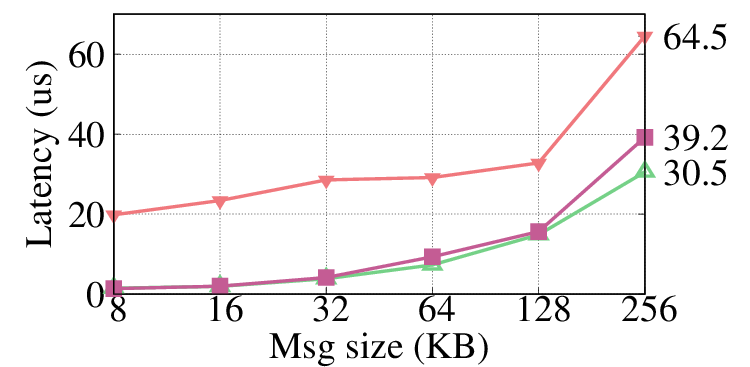}
  \footnotesize 
  \textbf{(b) Latency (large msg).}
  \end{minipage}

  \begin{minipage}[t]{0.48\linewidth}
  \centering 
  \includegraphics[width=0.99\textwidth]{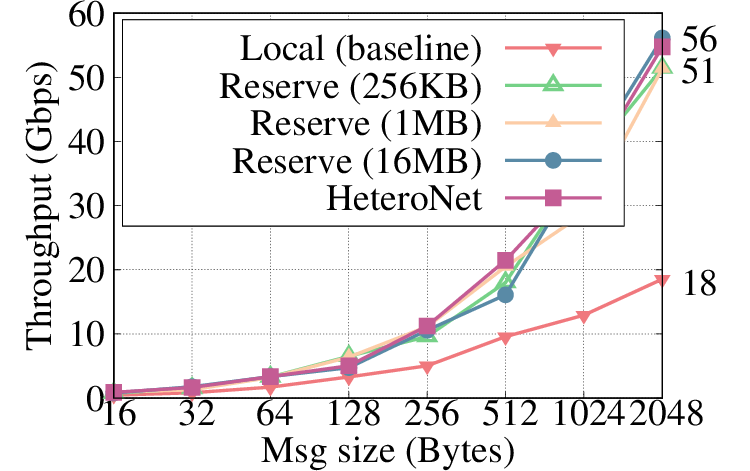}
  \footnotesize 
  \textbf{(c) Throughput (Shm, small).}
  \end{minipage}
\begin{minipage}[t]{0.48\linewidth}
  \centering 
  \includegraphics[width=0.99\textwidth]{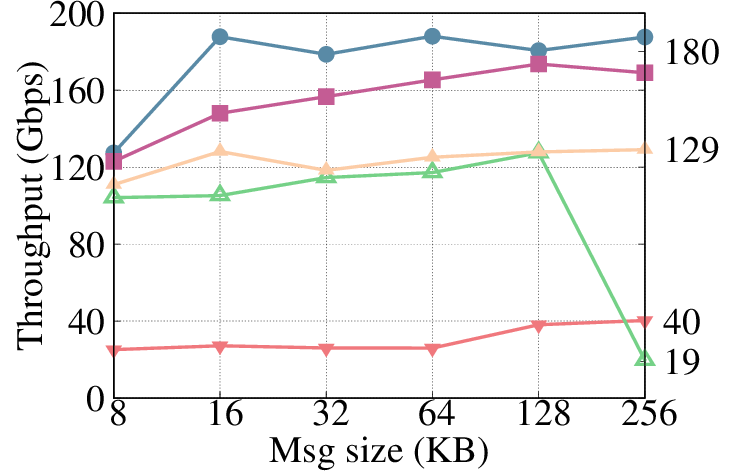}
  \footnotesize 
  \textbf{(d) Throughput (Shm, large).}
  \end{minipage}

  \begin{minipage}[t]{0.48\linewidth}
  \centering 
  \includegraphics[width=0.99\textwidth]{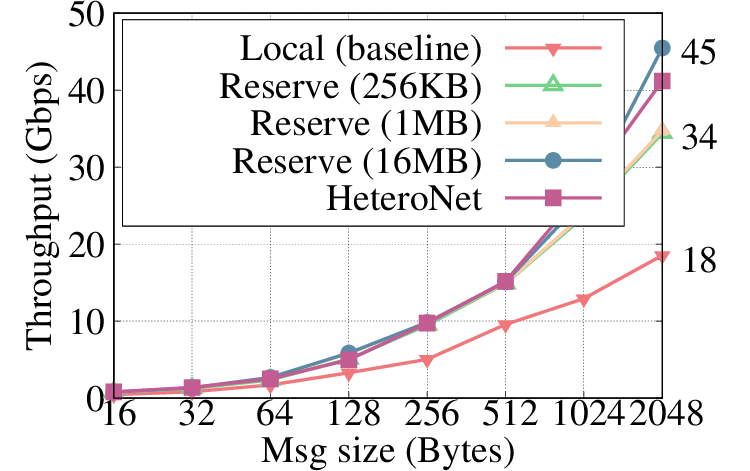}
  \footnotesize 
  \textbf{(e) Throughput (CXL, small).}
  \end{minipage}
\begin{minipage}[t]{0.48\linewidth}
  \centering 
  \includegraphics[width=0.99\textwidth]{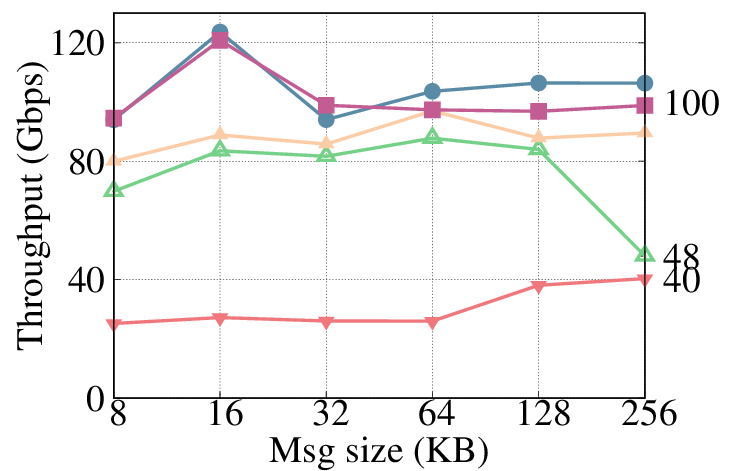}
  \footnotesize 
  \textbf{(f) Throughput (CXL, large).}
  \end{minipage}
	\caption{\textbf{Network performance.}}
  \label{fig:heterosocket-micro}
\end{figure}

\subsection{\gsocket Performance}
\label{subs:eval-socket-microbench}

We conduct an evaluation of network performance with \gsocket. 
We utilize two test cases, lat\_tcp and bw\_tcp from LMBench~\cite{mcvoy1996lmbench}, to evaluate network latency and throughput. 
We compare three systems:
(1) Baseline is the network performance in a single PU using Linux network,
(2) Reserve is a version of \os that reserves a specific GShm for connection, e.g., 1MB means the connection will reserve two 1MB buffers for use (two directions),
and (3) \os (128KB for local records and 128MB for shared arena).

\myparagraph{Latency.}
Fig.\ref{fig:heterosocket-micro}-a/b shows the results of latency for baseline and \os (local using shared memory and XPU using CXL).
Compared with baseline, \os can achieve 10.1x (2KB) -- 31.9x (16B) better latency for small msgs (in the same PU), and 2.1x (256KB) -- 13.2x (8KB) for large msgs.
For cross-PU cases, CXL-based \os can achieve comparable latency as Shm, and two orders of magnitude better over Cross-PU network baseline (not shown for space reason).
UNAPI takes about 0.51us for cross-PU notification with CXL, and needs additional costs to handle faults and upcall.

\myparagraph{Throughput.}
Fig.\ref{fig:heterosocket-micro}-c/d shows the results of throughput for local-PU (\os using Shm)
and e/f for cross-PU (\os using CXL, still use local-PU results for baseline because cross-PU network is significantly slower).
We also present the results of \os that utilizes fixed reserved buffer (instead of local-records/shared arena design). 
Compared with baseline, \os improves throughput by 1.9x--2.9x for small msgs and 4.8x--6.3x for large msgs on single-PU.
Cross-PU \os (CXL) outperforms single-PU baseline by 1.4x--4.4x.

\myparagraph{Resource costs.}
Assume 1,000 connections in Fig.\ref{fig:heterosocket-micro}-c/d/e/f.
Reserving 16MB will require 16GB pinned resource for communication, which is a non-trivial burden.
Instead, 1MB (1GB total) and 256KB (256MB total, shared by 1,000 connections) can achieve good throughput for small msgs, but show significant slowdown for larger msg sizes ($>$2x worse for 256KB).
\os uses 128KB private records and 128MB shared arena (256MB total), but achieves almost the same good performance as reserving 16MB --- 64x memory saving in this case.
Our practice in real cloud is to reserve 1MB for RDMA (per-connection) and 1MB for traditional Linux TCP,
therefore, the improvement is admirable.

\newcommand{\cmark}{\color[HTML]{32CB00}{\ding{51}}}
\newcommand{\xmark}{\color[HTML]{FE0000}{\ding{55}}}

\begin{table}[]
\setlength{\abovecaptionskip}{0pt}
\centering
\footnotesize
\caption{\textbf{LoC of supported applications.}
\textit{\os can support real-world applications with significant higher Line-of-Code (LoC). {\cmark} denotes supported by Molecule or \os, while {\xmark} signifies unsupported. CNN image is from FunctionBench, while Alexa is from ServerlessBench.}}
\begin{tabular}{@{}lcccc@{}}
\toprule
\textbf{Applications}    & \textbf{Version}                                                     & \textbf{LoC} & \textbf{Molecule} & \textbf{\os} \\ \midrule
Python3  & 3.8.10                                                               & 1,085,262    & \xmark                 & \cmark                   \\
Envoy                    & \#330dbdc8f9                                                         & 764,457      & \xmark                 & \cmark                   \\ \midrule
CNN image                & \#296f5f2  & 135          & \cmark                 & \cmark                   \\
Alexa smarthome          & \#207b345 & 1,804        & \cmark                 & \cmark                   \\ \bottomrule
\end{tabular} \\[-10pt]
\label{t:eval-efforts}
\end{table}

\subsection{Compatibility}
We analyze the Line-of-Code (LoC) for two of the most intricate applications supported by \sys and Molecule~\cite{10.1145/3503222.3507732},
as shown in Table~\ref{t:eval-efforts}. 
Although frameworks like Molecule utilizes new abstractions to sucessfully support serverless, it is very difficult to support commodity cloud-native apps like Python-based services and Envoy~\cite{envoy} (widely used in service mesh systems).
\sys (with \os) can support commodity apps with significantly higher LoC on CPU-DPU computers,
which is a new milestone.

\myparagraph{Supporting other heterogeneous devices.}
\os is general to other general-purpose heterogeneous devices.
However, for domain-specific accelerators like FPGA and GPU, applications usually do not depend on APIs like socket.
In these cases, we can directly use low-level abstraction, e.g., DMA, for efficient communication.
It is also possible to extend \os with related systems like GPUnet~\cite{kim2014gpunet}, Lynx~\cite{tork2020lynx}, and Coyote~\cite{korolija2020abstractions} in the future.

\myparagraph{Supporting other systems.}
We implement our prototype based on Kubernetes because it is the most widely-used cloud-native platform today~\cite{cncf-landscape}.
Nevertheless, the idea of dynamic split and the techniques (i.e., \giptable and \gsocket) are general.
The implementation can also be applied to other cases besides cloud-native apps.

\myparagraph{Limitations.}
Our prototype supports \gsocket for apps that dynamically linked with libc.
Statically pre-compiled apps are not supported (we can support static compiled apps with re-compilation).
\os mainly focuses on apps cooperate with network.
Prior work~\cite{10.1145/3503222.3507732} has supported the case of IPC, which will be integrated.
Offloading more fine-grained granularity, e.g., threads, is still an open challenge for future work.
More policies and benefits of XPU, e.g., energy-efficiency~\cite{234944}, are valuable to explore, but orthogonal to this work which enables the dynamic split \emph{mechanism}.

\section{Related Work}
\label{s:relwk}

\myparagraph{System supports for heterogeneous computers.}
There are many related research efforts~\cite{phothilimthana2018floem, seshadri2014willow, gu2016biscuit, baumann2009multikernel, barbalace2015popcorn, gamsa1999tornado, silberstein2013gpufs, kim2014gpunet, silberstein2017omnix, tork2020lynx, brokhman2019gaia, shan2018legoos, Asmussen:2016:MHC:2872362.2872371, nider2021last, pemberton2021restless, 10.1145/3458336.3465273, liu2019offloading, cho2020flick}.
Distributed OSes~\cite{baumann2009multikernel, silberstein2017omnix, barbalace2015popcorn} are proposed to run a single OS (with multi-kernels) to manage heterogeneous devices.
Some systems use smart devices to offload specific tasks~\cite{phothilimthana2018floem, seshadri2014willow, gu2016biscuit, kim2014gpunet, silberstein2013gpufs}.
Floem~\cite{phothilimthana2018floem} proposes programming abstractions for NIC-accelerated applications.
LeapIO~\cite{li2020leapio} leverages co-processors to offload cloud storage services.
E3~\cite{234944} is a microservice execution platform for SmartNIC-accelerated servers, which can significantly improve the energy-efficiency.
\os is the first work utilizing network abstraction to achieve dynamic split.
Our experience on \sys has proven the feasibility and efficiency, which is a new breakthrough.


\myparagraph{Optimizing cross-PU communication.}
Prior efforts~\cite{10.1145/2934872.2934897, 10.1145/2872362.2872401, 10.1145/3230543.3230560, 234944} have explored PCIe/DMA for efficient smartNIC-host communication.
GPUnet~\cite{kim2014gpunet} proposes a socket abstraction for GPU programs.
Solros~\cite{min2018s} and SPIN~\cite{bergman2017spin} utilizes P2P DMA to optimize communication.
Pond~\cite{10.1145/3575693.3578835} uses CXL for memory pooling.
CXL-Shm~\cite{10.1145/3600006.3613135} utilizes CXL for efficient DSM and resolves partial failure with reference counting.
\os is the first to provide a network abstraction.

\myparagraph{Related network solutions.}
SocksDirect~\cite{10.1145/3341302.3342071} introduces a user-space socket system,
utilizing a user-mode monitor to manage resources and events.
However, it does not target XPU computers, and still relies on reserved memory buffer.
SMC-R~\cite{rfc7609} supports shared memory-based network over RDMA in kernel.
It can effectively manage resources and support commercial apps.
However, the kernel-centric design incurs costs about switching and copying between user and kernel modes.
IX~\cite{belay2014ix} proposes a dataplane OS that provides both high I/O performance and protection.
Other kernel-bypass networking libraries that also preserve the POSIX API, e.g., Arrakis~\cite{peter2015arrakis}, mTCP~\cite{10.5555/2616448.2616493}, F-stack~\cite{f-stack},
usually utilize efficient user-space TCP/IP implementations and DPDK.
MegaPipe~\cite{10.5555/2387880.2387894} and StackMap~\cite{196184} optimize network stack with zero-copy APIs.
\os follows the line and proposes speculative allocation and UNAPI to balance performance and resource efficiency in a heterogeneous computer.

\section{Conclusion}
\label{s:conclusion}

This paper presents \xpod with \os, a novel design that enables dynamic split of cloud-native apps on CPU-DPU computers.
\os includes \giptable for unified network namespace and \gsocket to achieve both kernel-bypass performance and kernel-assisted resource-efficiency.
We build \sys to show the benefits on representative cloud-native scenarios including serverless computing and microservices.
\os will be open-sourced.

\bibliographystyle{ACM-Reference-Format}
\bibliography{ref}

\end{document}